**Singlet state encoded magnetic resonance (SISTEM) spectroscopy**

Andrey N. Pravdivtsev,*[a] Frank D. Sönnichsen[b] and Jan-Bernd Hövener[a]

[a] Section Biomedical Imaging, Molecular Imaging North Competence Center (MOIN CC), Department of Radiology and Neuroradiology, University Medical Center Kiel, Kiel University (http://www.moincc.de/)
Am Botanischen Garten 14, 24118, Kiel, Germany
E-mails: andrey.pravdivtsev@rad.uni-kiel.de; jan.hoevener@rad.uni-kiel.de

[b] Otto Diels Institute for Organic Chemistry, Kiel University, Otto Hahn Platz 5, 24098, Kiel, Germany

**Abstract.** Magnetic resonance spectroscopy (MRS) allows the analysis of biochemical processes non invasively and in vivo. Still, its application in clinical diagnostics is rare. Routine MRS is limited to spatial, chemical and temporal resolutions of cubic centimetres, mM and minutes. In fact, the signal of many metabolites is strong enough for detection, but the resonances significantly overlap, exacerbating identification and quantification. In addition, the signals of water and lipids are much stronger and dominate the entire spectrum. To suppress the background and isolate selected signals, usually, relaxation times, J-coupling and chemical shifts are used. Here, we propose methods to isolate the signals of selected molecular groups within endogenous metabolites by using long-lived spin states (LLS). We exemplify the method by preparing the LLSs of coupled protons in the endogenous molecules N-acetyl-L-aspartic acid (NAA). First, we store polarization in long-lived, double spin states and then apply saturation pulses and double quantum filters to suppress background signals. We show that LLS can be used to selectively prepare and measure the signals of chosen metabolites or drugs in the presence of water, inhomogeneous field and highly concentrated fatty solutions. The pH measurement presented here is one of the possible applications.

## 1. Introduction

Magnetic resonance (MR) has found a multitude of applications in medical imaging, from anatomy to motion, function and metabolism.[1–3] Likely, one of the most promising, yet least delivering applications is in vivo MR spectroscopy (MRS). MRS provides a non-invasive window into the biochemistry of living organisms – basically a virtual biopsy. Alas, it does not live up to this promise, as MRS is rarely used in routine diagnostics.

The success of MRS is hampered by the interplay of two major issues: first, a comparably weak signal of the metabolites of interest, and second, confounding overlapping resonances and strong background signals mostly from water and lipids. As a result, most of the time, only highly concentrated metabolites are detected, such as the major brain metabolites N-acetyl-L-aspartic acid (NAA), creatine (Cr), choline (Cho), myo-Inositol (Myo), lactate and alanine. For example, up to 0.1 % of the brain tissue wet weight of mammals belongs to NAA[4] which is associated with neuronal integrity.[5,6]

$^2$H-enriched, nonradioactive molecules like $^2$H-glucose are used to obtain background-free spectra in vivo. Valuable insights into biochemistry were gained by this approach.[7,8] However, because of the limited amount of substrate that can be administered in conjunction with the low gyromagnetic ratio of $^2$H, the signal remains low and does not allow routine high-resolution imaging.

Hyperpolarization techniques boost the signal of selected, isotopically labelled metabolites. This way, the fate of a dedicated, polarized agent can be followed in vivo with increased spatial, chemical and temporal resolution.[9–11] For example, hyperpolarization allows imaging the distribution of the hyperpolarized agent, mapping of tissue pH or metabolic conversion.[12] These methods provide different information than conventional MRS without injections, where the stable metabolic equilibrium is measured.

In humans, hyperpolarization has shown great promise for imaging prostate cancer, brain cancer, monitoring therapy response, heart metabolism and lung imaging.[13–18] Still, the method is limited by the relatively short lifetime of the signal enhancement and thus short observation window, as well as the limited amount of hyperpolarized agent that can be injected. Additional requirements include a polarizer, specialized imaging sequences and an X-nuclei channel for the MR system.[19–22]

In fact, one may argue that the signals of many metabolites are already strong enough for many applications without further enhancement. Unfortunately, it is difficult to isolate the signal of an individual metabolite in the densely packed $^1$H spectrum of the human brain; the resonances of many interesting metabolites overlap or differ only by a few parts per million.

Thus, much inventiveness has gone into the development of suppression and spectral editing techniques,[6,23–26] exploiting differences in chemical shift, relaxation times or J-couplings.

The presence of numerous small molecules, "macromolecules" and fat greatly complicates the analysis of a spectrum; all aliphatic protons abundant in fat and small molecules occupy the same chemical shift region of 1-5 ppm. Therefore, usually, only the sharpest singlets or doublets signals of $CH_3$ protons of small molecules are prominent in $^1$H-MRS.[1,6,25,26]

In order to get more information from the spectra, the suppression of undesired chemicals using contrast-enhancing singlet states (SUCCESS) was proposed by DeVience et al.[27] Not unlike SUCCESS, here, we selectively prepare the signal of specific, endogenous metabolites by means of singlet-state encoded MR (SISTEM). We show that this method may serve purposes beyond background suppression e.g. for imaging pH.

Often, the singlet state of strongly coupled nuclei is much longer lived ($T_{LLS}$) than the corresponding longitudinal magnetization ($T_1$).[28–30] It was suggested to use such long-lived states (LLS) as a new MR contrast,[31] to measure slow diffusion[32] or chemical exchange.[33] But the long lifetime is not the only unique property of LLSs. Just as interesting is that broad-band decoupling or continuous-wave excitation (CW) preserve these states and even prolong their lifetime,[34] while "normal" magnetization is depleted.

## 2. Methods and Materials

**2.1. Chemistry**

N-acetyl-L-aspartic acid (NAA, Sigma-Aldrich, 00920, CAS: 997-55-7), DL-lactic acid (Sigma-Aldrich, 69785, CAS: 50-21-5), L-alanine (Sigma-Aldrich, A7469, CAS: 56-41-7), creatine monohydrate (Sigma-Aldrich, C3630, CAS: 6020-87-7), choline chloride (Sigma-Aldrich, C7017, CAS: 67-48-1), L-glutamic acid (Sigma-Aldrich, 49449, CAS: 56-86-0), myo-Inositol (Sigma-Aldrich, I7508, CAS: 87-89-8) were purchased and used without further purification.

**Model solution 1 (MS1)**: Each of the substrates listed above was dissolved in $D_2O$ (Deutero GmbH, 00506) to yield a concentration of 10 mM. pH was adjusted to the desired value by adding NaOD (Deutero GmbH, 03703) or DCl (Sigma Aldrich, 543047); pH-dependent NMR spectra of all substrates are given in **SI**. Note that only NAA is discussed in the main text.

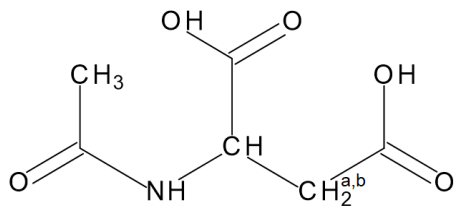

**Scheme 1. Chemical structure of N-acetyl-L-aspartic acid (NAA).** The methyl protons (a, b) were used for SISTEM. The protons have J-coupling constant of 16 -17 Hz with a chemical shift of 2.5 – 3 ppm depending on pH (**SI**).

**Model solution 2 (MS2):** a solution of 10 mM NAA in 300 µl $D_2O$ was mixed with 300 µl food-grade dairy cream (30% fat concentration).

**Model solution 3 (MS3):** 100 mM NAA were dissolved in 1500 µl deionized $H_2O$.

### 2.2. NMR and MRI

All high-resolution NMR spectra were acquired on a 600 MHz spectrometer (Bruker Avance II) with a cryogenically cooled probe (TCI) and 5 mm NMR tubes (Wilmad).

Nonlocalized spectra were acquired with a 7 T MRI with a 30 cm diameter of the inner bore (Bruker 7/30 ClinScan), equipped with a [1]H transmit-receive volume coil with an inner diameter of 8 cm for excitation and a single loop surface coil with an inner diameter of 0.9 cm for detection.[35] **MS3** was filled into a 1.5 ml container (Eppendorf vial) and placed in the pick-up coil at the isocenter of the magnet.

Routine nonlocalized spectroscopy and inversion recovery sequence were used; SISTEM-1 and 2 were implemented using the manufacturer's software (TopSpin 3.2 or Siemens VB15, IDEA 1.5b1.63). The same software was used to process data. Only square RF-pulses and trapezoidal gradients were used.

### 2.3. Quantification methods

The signals were quantified by integrating phased (or amplitude spectra, if explicitly mentioned).

A mono-exponential recovery function was fitted to integrals to extract the $T_1$-relaxation times.

To measure the lifetime of LLS, $T_{LLS}$, SISTEM-1 was repeated for various settings of $\tau_d$ (**Figure 2A**). A two exponential decay function was fitted to the data, and the longest time was attributed to $T_{LLS}$.

The signal in SISTEM-1 experiment as a function of $\tau_1$-interval was fitted with the function:

$$F_1 = \left[\sin\left(2\pi J_{CH_2^a}^{CH_2^b}\tau_1\right)\cos(\pi\Delta J\tau_1)\cos(\pi\Sigma J\tau_1)\right]^2 \qquad (1)$$

where $J_{CH_2^a}^{CH_2^b}$, $J_{CH_2^a}^{CH}$ and $J_{CH_2^b}^{CH}$ are corresponding J-coupling constants of 3 spin-½ system of NAA ($CH_2$ and CH) and $\Delta J = J_{CH_2^a}^{CH} - J_{CH_2^b}^{CH}$, $\Sigma J = J_{CH_2^a}^{CH} + J_{CH_2^b}^{CH}$.

This equation was obtained by using product operators[36] in the same way as was done for a 2-spin system.[33,37] The cosines are a result of the J-coupling interaction of $CH_2$ protons with a CH. The square of all component is a result of magnetization transfer to singlet state and backwords.

### 2.4. Pulse sequence

In general, SISTEM is composed of five steps with different functions (**Figure 1**), some of which may occur at the same time.

<u>Step 1</u>: In the first stage, thermal spin magnetization is transferred to the population of a singlet spin state and zero-quantum coherences (ZQCs). To reach this goal, several methods were proposed.[38–41] We chose the method proposed by Sarkar et al.[33] (Sarkar-II) because it is independent of frequency offsets and uses only hard RF-pulses and gradients. To suppress high order quantum coherences, we added a spoiler gradient after this stage (**Figure 2A, 4A**).

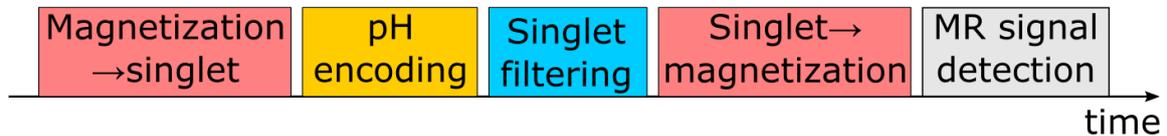

**Figure 1. Suggested 5-step SISTEM pulse sequence:** (1) magnetization of ¹H nuclei is transferred to their LLSs (and zero-quantum coherences); (2) encoding a feature (here: chemical shift difference that can be correlated with the pH value, **Figures 4, 5**); (3) background suppression: only singlet spin states (or with zero-quantum coherences) pass through the filter; (4) conversion of singlet state into observable magnetization and (5), MR signal detection.

<u>Step 2</u>: The second stage is used to encode a dedicated feature (e.g. chemical exchange,[33] diffusion[42] or pH). Here, we encode the chemical shift difference of NAA-$CH_2$ protons by introducing a free evolution interval $\tau_2$ (**Figures 2A, 4A**). During this stage, in-phase and out-of-phase ZQCs mutually alternate.[33]

<u>Step 3</u>: The goal of the third stage is to suppress all non-singlet states. We implemented two variants: strong broadband decoupling on the NMR device (**SISTEM-1, Figure 2A**), and Only-Parahydrogen SpectroscopY with double quantum coherence filter (OPSYd) on the MRI system (**SISTEM-2, Figure 4A**).[35,43] Note that the deposition of r.f. energy by OPSYd is much lower than that of decoupling.

Step 4: During the fourth stage, LLSs and ZQCs are transferred back to observable magnetization. For SISTEM-1, we used the second block of the Sarkar-II sequence (out of phase echo block, **Figure 2A**). For SISTEM-2, this function is accomplished with OPSYd (**Figure 4A**).

Step 5: During the final stage, MR signal is acquired e.g. by pulse-acquisition experiments or, possibly, imaging.

## 3. Results

### 3.1. SISTEM-1 on a high-resolution NMR spectrometer

First, we used the NMR spectrometer to optimize SISTEM-1 for preparing the NAA-CH$_2$ resonance. (SARKAR-II,[33] **Figure 2A, MS1**). $^1$H spectra were acquired for 13 different settings of $\tau_1$, and eq. 1 was fitted to the data. An optimal $\tau_1 \approx$ 12 ms was determined (**Figure 2B**).

It should be noted that SISTEM-I was designed with a 2-spin-½ system in mind. NAA, however, is effectively a 3 spin-½ system (CH-CH$_2$, **Scheme 1**), therefore the efficiency of the sequence was reduced. Still, the LLS was successfully populated, and the lifetime of T$_{LLS}$ = 6.5 s was almost 6 times longer than T$_1 \approx$ 1 s (**Figure 2B,C,** at 14.1 T with 2.5 kHz WALTZ-16 decoupling[44]).

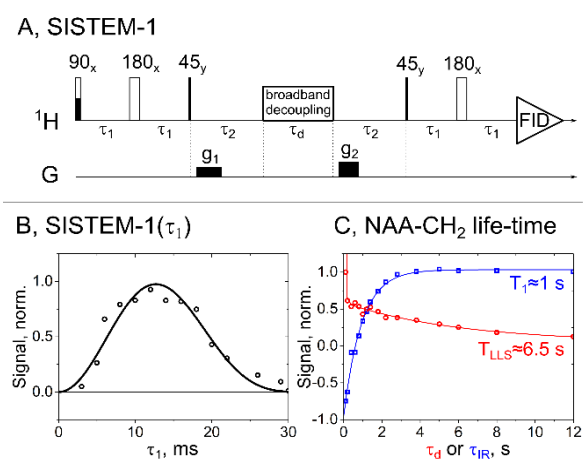

**Figure 2. Scheme (A) and optimization (B) of SISTEM-1, as well as the measurement of T$_{LLS}$ and T$_1$ for NAA-CH$_2$ (C).** To optimize SISTEM-1, the NAA-CH$_2$ signal was acquired as a function of $\tau_1$ and eq. 1 was fitted to the data (B, $\tau_d$ = 1 s). The lifetime of LLS was measured to T$_{LLS} \approx$ 6.5 s by varying $\tau_d$ (C, red $with\ \tau_1$ = 8 ms), and fitting a two exponential decay function to the data. T$_1$ of NAA-CH$_2$ was measured by an inversion recovery (IR) experiment, yielding T$_1 \approx$ 1 s (C, blue). Parameters: WALTZ-16 decoupling with 2.5 kHz RF-field amplitude at 600 MHz. No phase cycling was used.

Next, we used a 1:1 mixture of water and dairy cream with 30% fat (**MS2**, **Figure 3B, C**). To evaluate the performance of the method in a field with poor homogeneity, we refrained from shimming. The resulting linewidth was irregular with a full width at half maximum of ≈ 30 Hz, resembling in vivo conditions.

As expected, the spectrum of a standard, 90° pulse-acquisition experiment (without any suppression techniques) was dominated by fat and water resonances, while only very little NAA was apparent (**Figure 3B**). Using SISTEM-1, however, the fat and water signals were strongly suppressed, allowing to increase the receiver gain (rg) 320 times (linear, **Figure 3C**).

The SISTEM-1 spectrum showed well-resolved resonances of NAA-$CH_2$, but no signal from the NAA-$CH_3$ group, which normally dominates the spectrum. In addition, some water signal and some residues of lipid-$(CH_2)_n$ were found at ≈ 1.3 ppm. The signals in the range 3.5 – 4.5 ppm were attributed to lactose resonances, which is expected to be abundant in a dairy cream.

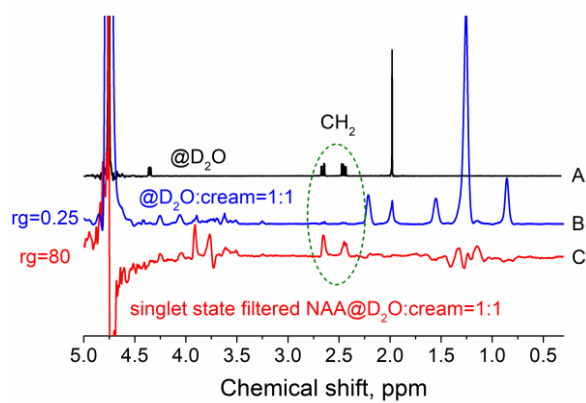

**Figure 3. $^1$H NMR spectra of NAA in $D_2O$ (MS1) and homogeneous field (A), NAA in aqueous-cream solution (MS2) with poor field homogeneity acquired by standard NMR (B) and edited by SISTEM-1 (C).** Water (4.7 ppm) and lipid resonances (0.9 – 2.2 ppm range)[45] dominated the routine NMR spectrum (B) but were effectively suppressed by SISTEM-1 (C), allowing to increase the receiver gain (rg) by 320 folds. The NAA-$CH_2$ resonance was well prepared by SISTEM-1 (dashed oval). Interestingly, two other resonances at ≈ 3.8 ppm were also prepared by SISTEM-1. These chemical shifts coincide well with those of lactose.[46] Parameters of SISTEM-1: $\tau_1 = 8$ ms, $\tau_d = 1$ s, number of scans was 8, no phase cycling, WALTZ-16 decoupling with 2.5 kHz RF- amplitude.

### 3.2. SISTEM-2 on a 7 T small animal MRI system

For in vivo applications, care must be taken when using decoupling because much energy may be deposited in the tissue. To circumvent this issue, we propose to use OPSYd instead of decoupling (SISTEM-2, **Figure 4A**). Note that in this case, an additional magnetization to singlet transfer stage is no longer required.

We implemented SISTEM-2 on a preclinical 7 T MRI. Again, the spectrum of a standard pulse-acquisition method was dominated by water signal three orders of magnitude larger than that of NAA (**MS3**). Water and NAA-CH$_3$ signals were strongly suppressed by SISTEM-2, while NAA-CH$_2$ signal was retained (**Figure 4B**, **C**).

Selectively preparing specific resonances and suppressing background opens the door for new applications. Here, we use it to encode the chemical shift difference of NAA-CH$_2$ protons ($\Delta\delta_{CH_2}$) and thus the pH of the solution (see below). Consider the following: the zero-quantum coherences evolve during the time period $\tau_2$ with the frequency equal to $\Delta\delta_{CH_2}$ (**Figure 4D**). Thus, acquiring SISTEM-2 spectra with different $\tau_2$ should allow to precisely measure $\Delta\delta_{CH_2}$. Experimentally observed additional modulation of SISTEM-2 signal as a function of time (**Figure 4E**) is caused by a spin-spin interaction with the third CH-proton[47,48] and leads to a doublet in the spectrum (**Figure 4E**). The middle point of this doublet is a chemical shift difference $\Delta\delta_{CH_2}$ = 53.4 Hz at 7 T or 0.178 ppm. Note that it is impossible to determine such splitting in a simple pulse-acquisition experiment with a common in MRI field homogeneity worse than 30 Hz.

Now that the chemical shift difference of NAA-CH$_2$ can be measured much more precisely than the poor homogeneity of the magnetic field would normally allow, we can use this information to assign pH of the sample. For this purpose, we acquired 7 high-resolution NMR reference spectra of NAA at pH 3 – 10. $\Delta\delta_{CH_2}$ was found to collapse for low pH approaching 2, and to reach a maximum for a pH of 7 or more (≈ 0.2 ppm, **Figure 5**). According to this data, the measured $\Delta\delta_{CH_2}$ = 53.4 Hz at 7 T corresponds to a pH of 5.25±0.1, which compares well with the value measured with an electrode.

Because $\Delta\delta_{CH_2}$ within one and the same molecule is used as pH meter, the method does not require any external reference, nor is any correction of magnetic field homogeneity or susceptibility needed.[49,50] Unfortunately, though, for NAA, most of the $\Delta\delta_{CH_2}$ variation is just below the most interesting physiological range of pH. Still, other molecules with more appropriate properties may be identified.

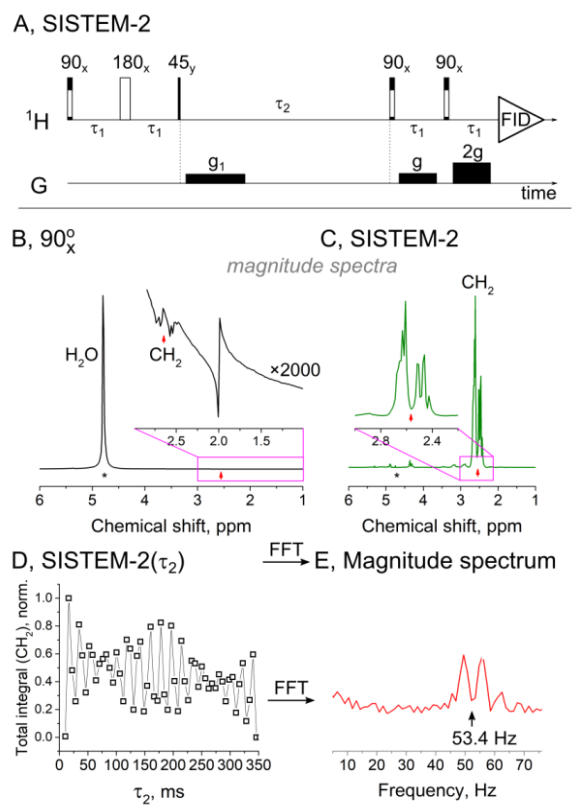

**Figure 4. SISTEM-2 on a 7 T MRI and $\Delta\delta_{CH_2}$ evaluation.** Schematic view of SISTEM-2 (A). In comparison to a pulse-acquisition magnitude spectrum (B), the water signal of **MS3** was strongly suppressed by SISTEM-2, while the NAA-CH$_2$ signal was maintained (C, magnitude). The chemical shift difference $\Delta\delta_{CH_2}$ was determined by Fourier transforming the summed amplitudes of the SISTEM-2 signal acquired for different $\tau_2$ (D). A well-resolved doublet at $\Delta\delta_{CH_2}$ = 53.4 Hz or ≈ 0.18 ppm was obtained (E) despite a poor field homogeneity (worse than 30 Hz linewidth in A). Inserts show NAA-CH$_2$ resonances (B, C).

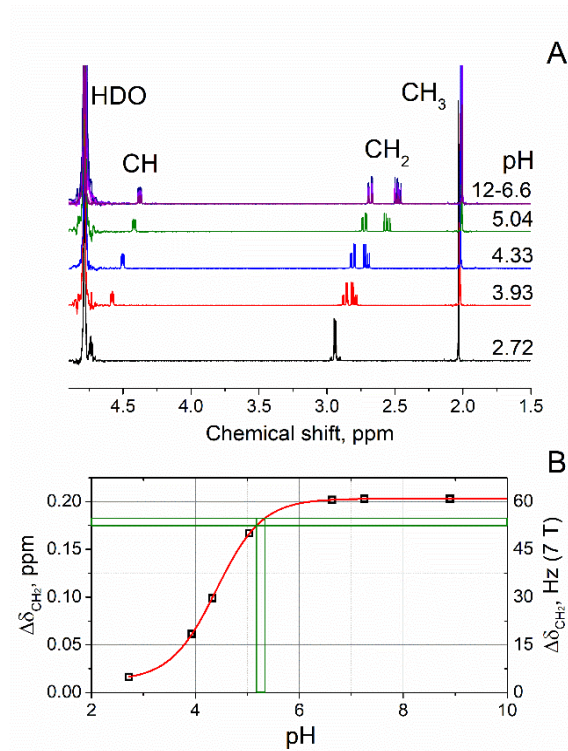

**Figure 5. SISTEM-2 allows using $\Delta\delta_{CH_2}$ as pH meter.** $^1$H-NMR spectra (A, **MS1**) and chemical shift difference (in ppm or in Hz at 7 T) of NAA-CH$_2$ protons as a function of pH (B). NMR spectra were acquired at 600 MHz, but the relative difference (ppm) is independent of the field strength. The modified Hill equation[51] was fitted to the $\Delta\delta_{CH_2}$ curve (red line). The intersection of green boxes shows the measured value of the chemical shift difference and the corresponding pH value.

## 4. Discussion

Among the multitude of applications suggested for LLS, we showed here that LLS can be used to selectively prepare the signals of dedicated metabolites in the presence of water, inhomogeneous magnetic field and highly concentrated fatty solutions. Magnetization prepared with SISTEM is easier to quantify because the background is greatly suppressed.

One application of this technique may be the measurement of endogenous brain metabolites, either of a whole-brain at once[6] or as a part of localized spectroscopy methods. Another application may be to image the biodistribution of drugs such as ethosuximide (ETX, Zarontin).[52] ETX is a medication used to treat absence seizures. Simulations suggest that SISTEM will work on ETX just as well as on NAA.

The measurement of pH values may be another promising application. Use of ZQCs for pH encoding is advantageous because no external reference is needed and this value is not affected by magnetic field inhomogeneity. The sequence is not limited to CH$_2$ groups only. For example, fragments such as CH-CH$_3$ (e.g. lactate, alanine) or other groups of molecules of interest can be selected as well.[27]

While the conditions investigated here were used to approach the situation in vivo, obviously, only in vitro experiments were presented here. If the method has any value for in vivo biomedical applications has yet to be shown. Still, these first results are promising.

## 5. Conclusion

We showed that LLS prepared by SISTEM can be used to selectively prepare and measure the signals of chosen metabolites or drugs in the presence of water, inhomogeneous field and highly concentrated fatty solutions. Very strong suppression of unwanted signals was observed on an NMR and MRI system, and the chemical shift difference of the NAA-CH$_2$ protons was used for measuring pH. In vivo tests are required to see if this method has a biomedical relevance.

**Supporting materials**

NMR spectra of all substrates listed in the methods and chemical shifts and J-coupling values for NAA as a function of pH.

**ORCID**

Andrey Pravdivtsev https://orcid.org/0000-0002-8763-617X

F. D. Sönnichsen https://orcid.org/0000-0002-4539-3755

Jan-Bernd Hövener https://orcid.org/0000-0001-7255-7252

**Acknowledgements**

We acknowledge support by the Emmy Noether Program "metabolic and molecular MR" (HO 4604/2-2), the research training circle "materials for brain" (GRK 2154/1-2019), DFG - RFBR grant (HO 4604/3-1, № 19-53-12013), the German Federal Ministry of Education and Research (BMBF) within the framework of the e:Med research and funding concept (01ZX1915C), cluster of Excellence "precision medicine in inflammation" (PMI 1267). Kiel University and the Medical Faculty are acknowledged for supporting the Molecular Imaging North Competence Center (MOIN CC) as a core facility for imaging in vivo. MOIN CC was founded by a grant from the European Regional Development Fund (ERDF) and the Zukunftsprogramm Wirtschaft of Schleswig-Holstein (Project no. 122-09-053).

**Abbreviations**:

MRS – Magnetic resonance spectroscopy

LLS – long-lived spin states;

NAA – N-acetyl-L-aspartic acid;

SUCCESS – suppression of undesired chemicals using contrast-enhancing singlet states

SISTEM – singlet-state encoded magnetic resonance spectroscopy

CW – continuous-wave excitation

OPSYd – Only-Parahydrogen SpectroscopY with double quantum coherence filter

# Supporting materials

## Singlet state encoded magnetic resonance (SISTEM) spectroscopy

Andrey N. Pravdivtsev,*[a] Frank D. Sönnichsen[b] and Jan-Bernd Hövener[a]

[a] Section Biomedical Imaging, Molecular Imaging North Competence Center (MOIN CC), Department of Radiology and Neuroradiology, University Medical Center Kiel, Kiel University (http://www.moincc.de/)
Am Botanischen Garten 14, 24118, Kiel, Germany
E-mails: andrey.pravdivtsev@rad.uni-kiel.de; jan.hoevener@rad.uni-kiel.de

[b] Otto Diels Institute for Organic Chemistry, Kiel University, Otto Hahn Platz 5, 24098, Kiel, Germany

Table of content

1. N-acetyl-L-aspartic acid

**Table S1.** Chemical shifts, $\delta$, (in ppm) and J-coupling constants, $J$, (in Hz) of NAA protons as a function of pH (Figure 5). NMR parameters are constant at pH above ~6. NMR spectra acquired on a Bruker Avance II 600 MHz were analyzed.

| pH | $\delta(CH_3)$ | $\delta(CH_2^a)$ | $\delta(CH_2^b)$ | $\delta(CH)$ | $\delta(CH_2^b) - \delta(CH_2^a)$ | $J(CH_2)$ | $J(CH_2^a - CH)$ | $J(CH_2^a - CH)$ |
|---|---|---|---|---|---|---|---|---|
| 2.72 | 2.03306 | 2.93212 | 2.9485 | 4.73791 | 0.01638 | 17.2 | 6 | 6 |
| 3.93 | 2.02189 | 2.80271 | 2.86423 | 4.58047 | 0.06152 | 16.6 | 7.2 | 4.8 |
| 4.33 | 2.01658 | 2.71001 | 2.80899 | 4.504 | 0.09898 | 16.2 | 7.9 | 4.5 |
| 5.04 | 2.00821 | 2.55875 | 2.72563 | 4.4191 | 0.16688 | 15.9 | 9.5 | 3.9 |
| 6.63 | 2.0041 | 2.48064 | 2.68244 | 4.37907 | 0.2018 | 15.7 | 10.2 | 3.7 |
| 12 | 2.00422 | 2.47755 | 2.68048 | 4.37728 | 0.20293 | 15.7 | 10.2 | 3.7 |

**Table S2.** $T_1$ relaxation times of NAA in $D_2O$ at pH 6.

|  | $CH_3$ | $CH_2^a$ | $CH_2^b$ | $CH$ |
|---|---|---|---|---|
| $T_1$, s | 1.77 | 0.88 | 0.79 | 4.2 |

2. DL-Lactic acid

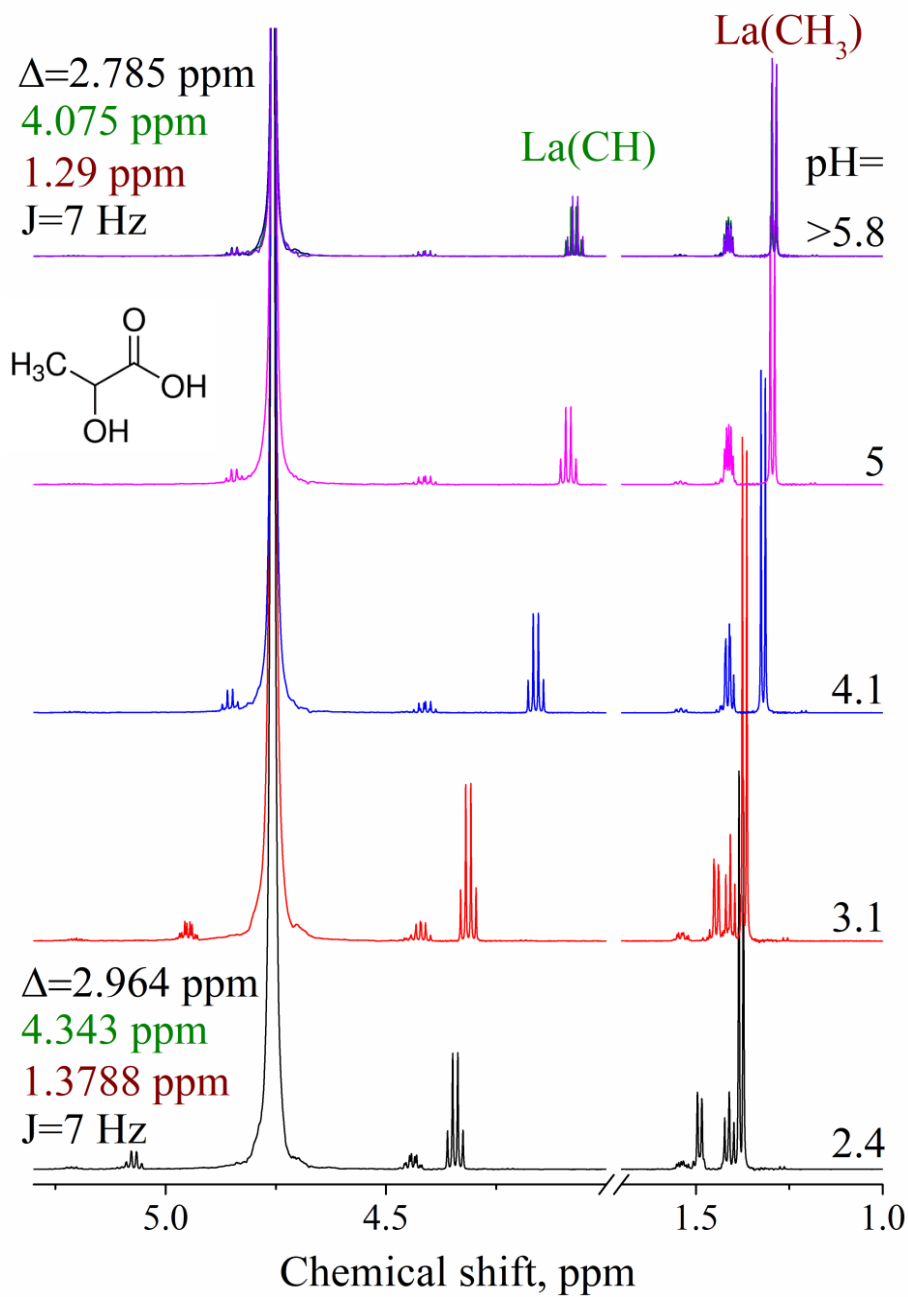

**Figure S1**. NMR spectra of 10 mM DL-Lactic acid (La, Sigma-Aldrich, 69785, CAS: 50-21-5) in D$_2$O at several pH values. Chemical shifts and J-coupling constants are indicated on figure. NMR spectra were acquired on a Bruker Avance II 600 MHz.

3. L-Alanine

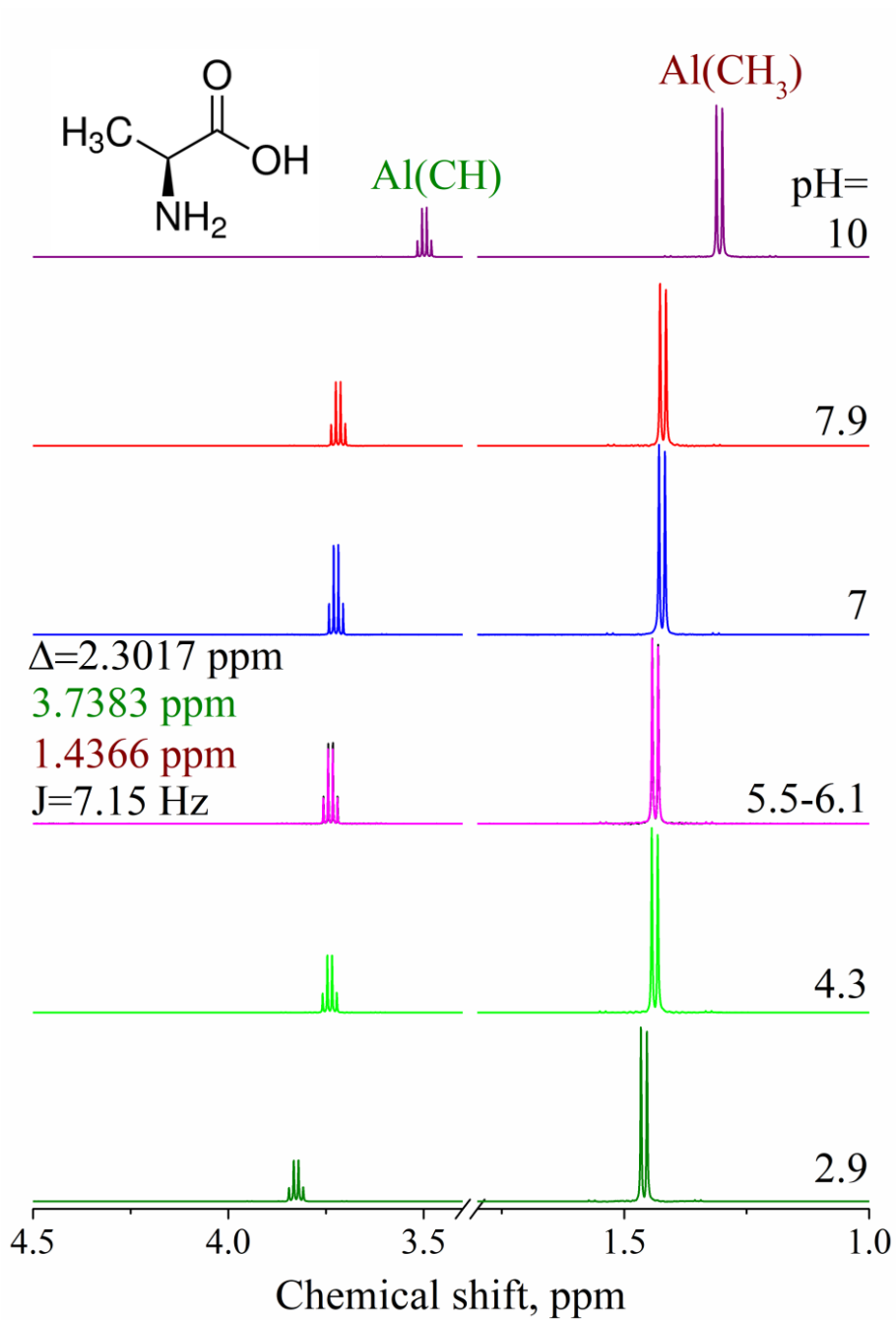

**Figure S2**. NMR spectra of 10 mM L-Alanine (Al, Sigma-Aldrich, A7469, CAS: 56-41-7) in $D_2O$ at several pH values. Chemical shifts and J-coupling constants are indicated on figure. NMR spectra were acquired on a Bruker Avance II 600 MHz.

4. Creatine monohydrate

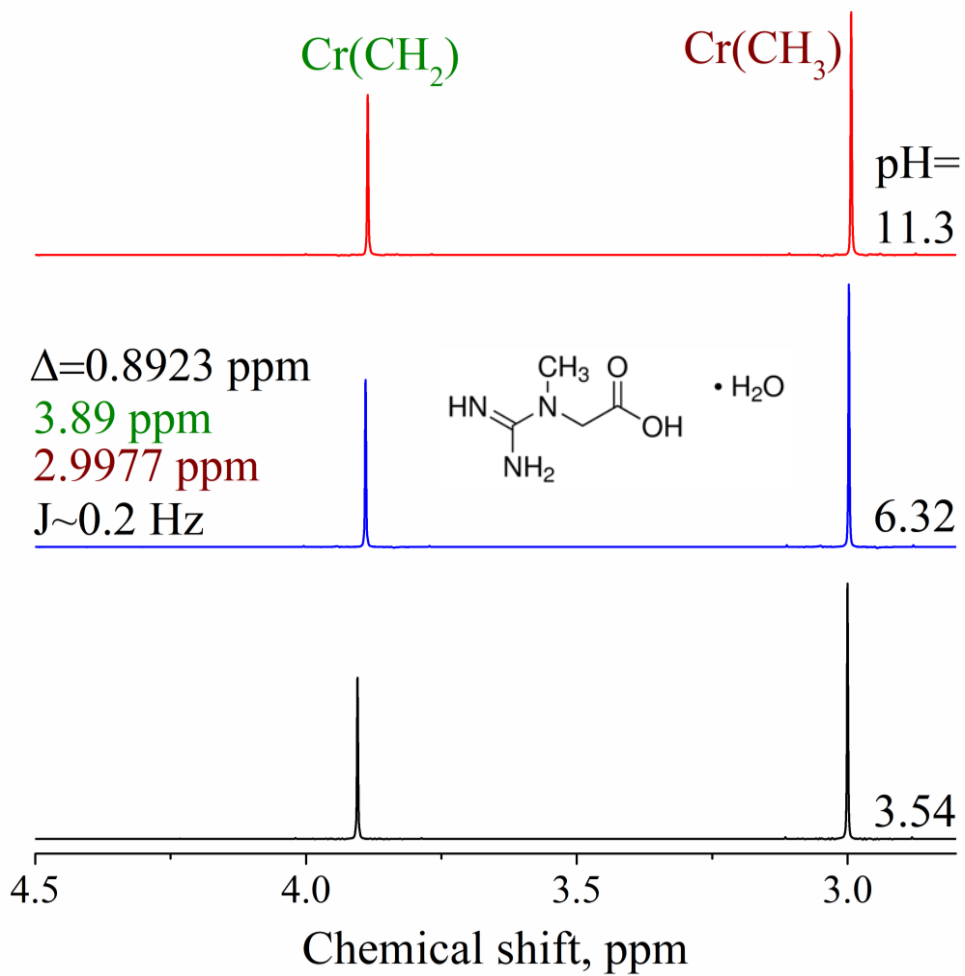

**Figure S3**. NMR spectra of 10 mM Creatine monohydrate ≥ 98% (Cr, Sigma-Aldrich, C3630, CAS: 6020-87-7) in D₂O at several pH values. Chemical shifts and J-coupling constants are indicated on figure. NMR spectra were acquired on a Bruker Avance II 600 MHz.

5. Choline chloride

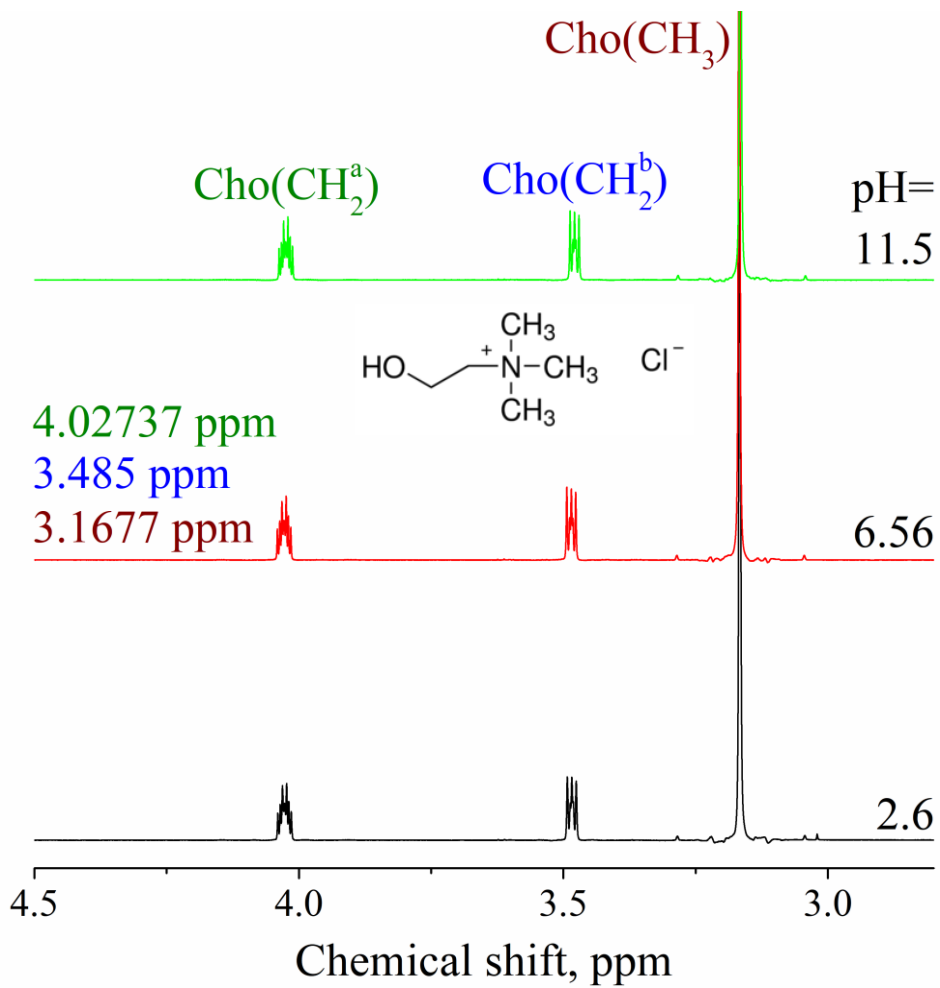

**Figure S4**. NMR spectra of 10 mM Choline chloride (Cho, Sigma-Aldrich, C7017, CAS: 67-48-1) in D$_2$O at several pH values. Chemical shifts and J-coupling constants are indicated on figure. NMR spectra were acquired on a Bruker Avance II 600 MHz.

6. L-Glutamic acid

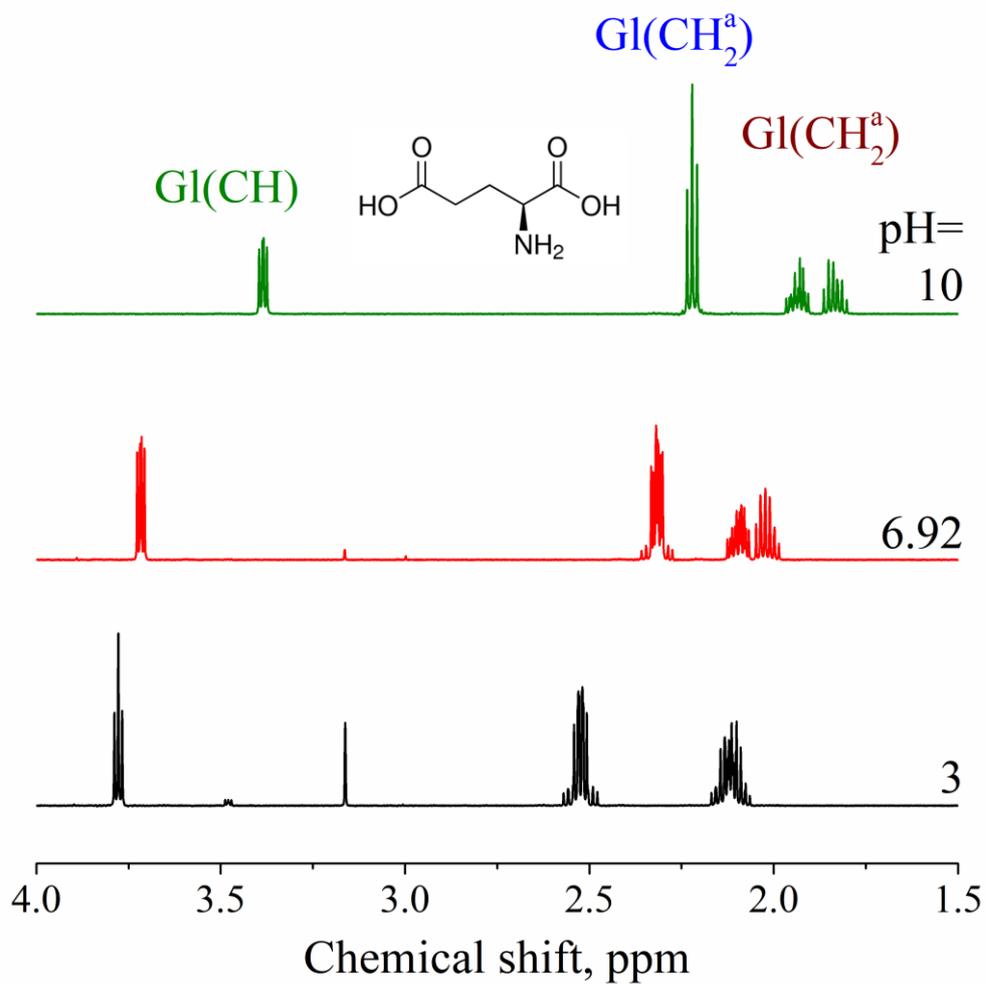

**Figure S5.** NMR spectra of 10 mM L-Glutamic acid (Gl, Sigma-Aldrich, 49449, CAS: 56-86-0) in D$_2$O at several pH values. NMR spectra were acquired on a Bruker Avance II 600 MHz.

7. Myo-Inositol

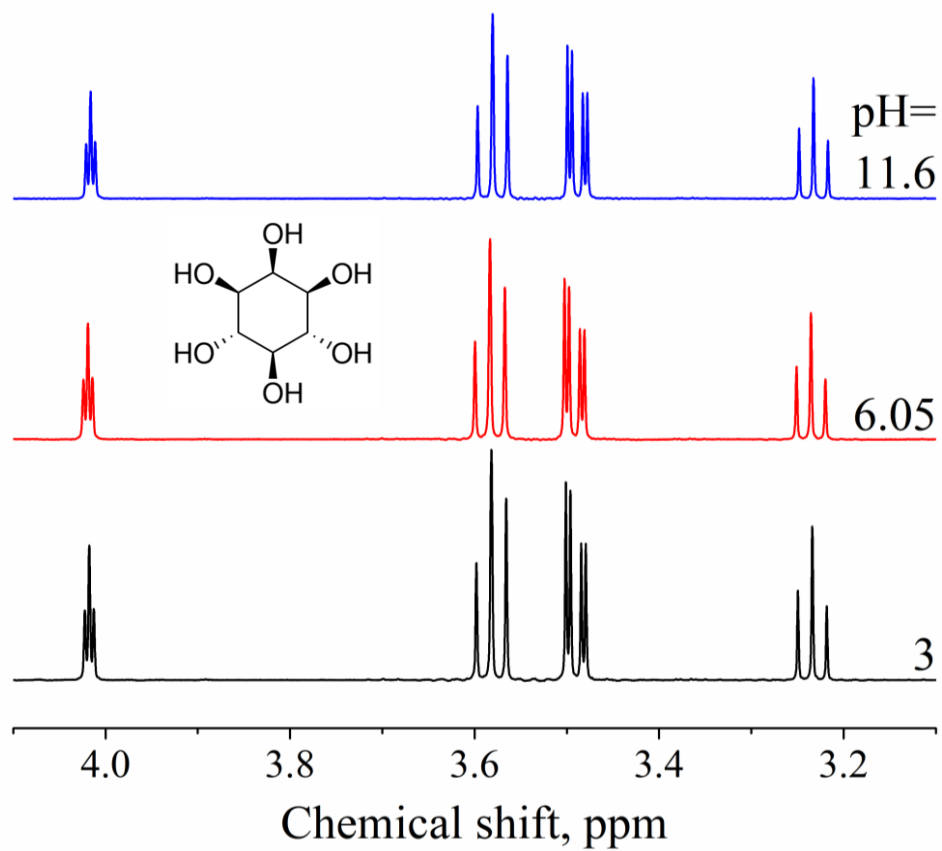

**Figure S6**. NMR spectra of 10 mM myo-Inositol (Sigma-Aldrich, I7508, CAS: 87-89-8) in $D_2O$ at several pH values. NMR spectra were acquired on a Bruker Avance II 600 MHz.